\documentclass[twocolumn,showpacs,preprintnumbers,superscriptaddress,amsmath,amssymb]{revtex4}
\usepackage{amssymb}		
\usepackage{amsmath}
\usepackage{graphicx}
\usepackage[normalem]{ulem}
\usepackage{multirow}
\usepackage{appendix}
\usepackage{CJK}
\usepackage[usenames]{color}
\usepackage{bm}
\usepackage[colorlinks,linkcolor=blue,urlcolor=blue,citecolor=blue]{hyperref}
\usepackage{booktabs} 	     
\usepackage{leftidx}

\begin{document}
\begin{CJK*} {UTF8} {gbsn}

\title{Signatures of $\alpha$ clustering in $^{16}$O  by using a multiphase transport model}

\author{Yi-An Li(李逸安)}

\affiliation{Shanghai Institute of Applied Physics, Chinese Academy of Sciences, Shanghai 201800, China}
\affiliation{Key Laboratory of Nuclear Physics and Ion-beam Application (MOE), Institute of Modern Physics, Fudan University, Shanghai 200433, China}
\affiliation{University of Chinese Academy of Sciences, Beijing 100049, China}

\author{Song Zhang(张松)}\thanks{Email: song\_zhang@fudan.edu.cn}
\affiliation{Key Laboratory of Nuclear Physics and Ion-beam Application (MOE), Institute of Modern Physics, Fudan University, Shanghai 200433, China}

\author{Yu-Gang Ma(马余刚)}\thanks{Email:  mayugang@fudan.edu.cn}
\affiliation{Key Laboratory of Nuclear Physics and Ion-beam Application (MOE), Institute of Modern Physics, Fudan University, Shanghai 200433, China}
\affiliation{Shanghai Institute of Applied Physics, Chinese Academy of Sciences, Shanghai 201800, China}

\date{\today}
\begin{abstract}
 $\alpha$-clustered structures in light nuclei could be studied through ``snapshots" taken by relativistic heavy-ion collisions. A multiphase transport (AMPT) model is employed to simulate the initial structure of collision nuclei  and the proceeding collisions at center of mass energy $\sqrt{s_{NN}}$ = 6.37 TeV.  This initial structure can finally be reflected in the subsequent observations, such as elliptic flow ($v_{2}$), triangular flow ($v_{3}$) and quadrangular flow ($v_4$). Three sets of the collision systems are chosen to illustrate system scan is a good way to identify the exotic $\alpha$-clustered nuclear structure, case I:  $\mathrm{^{16}O}$ nucleus (with or without $\alpha$-cluster) +  ordinary nuclei (always in Woods-Saxon distribution) in most central collisions,  case II:  $\mathrm{^{16}O}$ nucleus (with or without $\alpha$-cluster)  + $\mathrm{^{197}Au}$ nucleus collisions for centrality dependence, and  case III: symmetric collision systems (namely, $^{10}$B + $^{10}$B, $^{12}$C + $^{12}$C, $^{16}$O +  $^{16}$O (with or without $\alpha$-cluster),  $^{20}$Ne +  $^{20}$Ne,  and  $^{40}$Ca + $^{40}$Ca) in most central collisions. Our calculations propose that relativistic heavy-ion collision experiments at $\sqrt{s_{NN}}$ = 6.37 TeV are promised to distinguish the tetrahedron structure of $\mathrm{^{16}O}$ from the Woods-Saxon one and shed lights on the system scan projects in experiments. 	  
 \end{abstract}
\maketitle

	\section{Introduction}
	\par
	
Theoretical calculations predict that there is a new state of nuclear matter called quark-gluon plasma (QGP) created around a critical temperature $T_{c}$  $\approx170$ MeV~\cite{tc_1,tc_2}. In order to study the properties of this hot and dense matter, many experimental measurements have been performed at the Brookhaven National Laboratory - Relativistic Heavy Ion Collider (RHIC) ~\cite{ARSENE20051,BACK200528,ADAMS2005102,ADCOX2005184} as well as the CERN - Large Hadron Collider (LHC), followed by many theoretical studies~\cite{
Nagle,PBMPhysRep,ChenPhysRep,Bzdak,PBM,NatPhys,LuoNST,SongNST,MaYG,Tfree,Tfree2,Jpsi}. 
Observables such as collective flow~\cite{88014902,PhysRevLett112162301,PhysRevC88014904,PhysRevLett114252302}, Hanbury Brown-Twiss (HBT) correlation~\cite{PhysRevC92014904,adare2014beamenergy,PhysRevLett114142301}, chiral electric-magnetic effects~\cite{CME,CME2,PhysRevC91054901,PhysRevC94041901,WangFQ,XuZW,GaoJH} and fluctuation~\cite{PhysRevC92021901,LuoNST} have been proposed to extract information on the properties of hot dense matter formed in these collisions. These observables can inherit information of the collisions zone at very early stage (eg. for initial nuclear geometric asymmetry). From another perspective,  observables which are sensitive to initial geometry can give hints on the properties of initial state nuclei. $\alpha$-clustered structure as a specific phenomenon especially in light nuclei  where the mean field effect is not strong enough to break cluster structure is one of highly interesting topics in heavy-ion community, especially for well-known stable nuclei $^{12}$C and $^{16}$O. $\alpha$-cluster model was first proposed by Gamow and 
later on there are extensive studies, eg. 
~\cite{PhysRev521083,PhysRev54681,VONOERTZEN200643,PhysRevLett113032506,PhysRevC94014301,PhysRevC95034606,Huang_2017,Ye}. 
 However, definite $\alpha$-clustered configuration of carbon and oxygen still lacks sufficient experimental evidence. Traditionally, people believe that nuclear structure effect is significant only in low energy nuclear collisions. However, this kind of nuclear structure phenomenon can also be manifested through relativistic heavy-ion collisions. 
 Broniowski~\cite{Broniowski_2014,PhysRevC90064902} first proposed that this kind of nuclear structure phenomenon can be demonstrated through relativistic heavy-ion collisions. Observables such as harmonic flows can be measured with standard methods in relativistic heavy-ion collisions, offering a possibility to study low-energy nuclear structure phenomenon at relativistic colliders, such as RHIC and LHC~\cite{broniowski2015ultrarelativistic,PhysRevC97034912,PhysRevC100064912,PhysRevC95064904}.
Triggered by recent experimental measurements in small systems~\cite{PhysRevLett.121.222301,smallSystemQGPPHENIX,smallSystemSTAR,smallSystemCMS2013,smallSystemATLAS2016,smallSystemALICE2017}, some theoretical works of system scan were proposed at RHIC and LHC energies by using transport models or hydrodynamics models~\cite{PhysRevC99044904,PhysRevC101021901,PhysRevC100024904} in which collective phenomena with respect to geometric anisotropy were discussed therein. System dependence of heavy flavor production was also calculated in~\cite{rol2019size} by using Trento+v-USPhydro+DAB-MOD model.

This work aims at distinguishing the geometry structure of $\alpha$-clustered $^{16}$O nucleus in relativistic heavy-ion collisions. 
 From the results in this work, it is found that the participant multiplicity ($N_{part}$) dependence of the triangular flow over elliptic flow ($v_3/v_2$) can identify the exotic nuclear structure of $^{16}$O via system scan studies.
The rest of the paper is arranged as follows: In section~\ref{sec:model} a brief introduction of AMPT model and flow analysis methods are presented. The results and discussion are presented in section~\ref{sec:results}, and then a summary  is given.

\section{Model and methodology}
\label{sec:model}
		
A multi-phase transport model (AMPT) is developed to describe physics in relativistic heavy-ion collisions at RHIC~\cite{AMPT_origin} and is also suitable to reproduce some results at LHC by tuning some input parameters~\cite{AMPTGLM2016}, including pion-HBT correlations~\cite{AMPTHBT}, di-hadron azimuthal correlations~\cite{AMPTDiH,WangHai}, collective flows~\cite{STARFlowAMPT,AMPTFlowLHC,PhysRevC101021901} and strangeness production~\cite{NSTJinS,SciChinaJinS}. AMPT is a hybrid dynamic transport model, which consists of four main processes: (a) the initial conditions-HIJING model gives the spatial and momentum distributions of minijet partons and soft string excitations; (b) partonic cascade~\cite{ZPCModel}-interactions among partons are described by equations of motion for their Wigner distribution functions; (c) hadronization-conversion from the partonic to the hadronic matter; (d) hadronic interactions-based on the ART model~\cite{ARTModel}, including baryon-baryon, baryon-meson, and meson-meson elastic and inelastic scatterings. 
 
 The initial nucleon distribution in nuclei is configured in HIJING model~\cite{HIJING-1,HIJING-2} with either a pattern of Woods-Saxon distribution or  an exotic nucleon distribution is embedded to identify the $\alpha$-clustered structure of $^{16}$O through final state observables. For details, parameters of the tetrahedron structure $^{16}\mathrm{O}$ are inherited from an Extended Quantum Molecular Dynamics (EQMD) model~\cite{PhysRevLett113032506}, which is based on the quantum molecular dynamics (QMD) model. With the effective Pauli potential, EQMD model gives reasonable $\alpha$-cluster configurations for 4$N$ nuclei. 
 For four $\alpha$s in the tetrahedron structure, we put them at the vertexes with side length of 3.42 fm so that it gives a similar $RMS$-radius (2.699 $fm$) to the Woods-Saxon configuration (2.726 $fm$) as well as the experimental data (2.6991 $fm$)~\cite{ORMSRExp}, while nucleons inside each $\alpha$ are initialized by using the Woods-Saxon distribution introduced in HIJING model. 

Anisotropic flow is driven by the initial anisotropic density profile in high energy heavy ion collisions and is usually characterized by complex eccentricity coefficients ($\varepsilon _{n}$)  in the transverse plane as~\cite{PhysRevC86044908,vnen_beta_Ntrack201808,ZHANG2020135366} 
\begin{equation}
	\begin{aligned}
 \varepsilon _{n}e^{in\Phi_{n} }
		\equiv -\frac{\int d^{2} r_{\perp }r^{n}e^{in\varphi_{part}}\rho(r,\varphi_{part})}{\int d^{2} r_{\perp }r^{n}\rho(r,\varphi_{part})}
		\equiv -\frac{\left \langle r^{n}e^{in\phi } \right \rangle}{\left \langle r^{n} \right \rangle},
	\end{aligned}
	\label{eq:1}
\end{equation}
where $r=\sqrt{x^{2}+y^{2}}$ and $\varphi_{part}$ are the coordinate position and azimuthal angle of initial participant nucleons, $\rho(r,\varphi_{part})$ represents the density and $n$ is the order of the coefficients.
Definition here is necessary but not sufficient, if we take an elliptic Gaussian distribution
\begin{equation}
	\begin{aligned}
		\rho (x) = \frac{1}{2\pi \sigma _{x}\sigma _{y}}e^{-\frac{x^{2}}{2\sigma _{x}^{2}}-\frac{y^{2}}{2\sigma _{y}^{2}}}, 
		\label{eq:2}
	\end{aligned}
\end{equation}
as an example, we expect to get zero $\varepsilon _{4}$. However, if we calculate by the traditional definition Eq.~(\ref{eq:1}) we get nonzero value with an order of $\varepsilon_{2}^{2}$. Instead, we use the cumulants~\cite{PhysRevC86044908}
\begin{equation}
	\begin{aligned}
		C_{4}e^{i4\Phi _{4}}\equiv -\frac{1}{\langle r^{4}\rangle }[\langle r^{4}e^{i4\phi }\rangle-3\langle r^{2}e^{i2\phi } \rangle ^{2}]
		\label{eq:3}
	\end{aligned},
\end{equation}
\begin{equation}
\begin{aligned}
\varepsilon _{4}^{L} = \varepsilon _{4} + \frac{3\left \langle r^{2} \right \rangle^{2}}{\left \langle r^{4} \right \rangle}\varepsilon _{2}^{2}
\label{eq:4}
\end{aligned},
\end{equation}
rather than moments. Although the definition is still not mathematically sufficient, it is good enough to discuss current issues. And later we will see that flow calculations have the same problem.

\begin{figure*}[htb]
	\includegraphics[angle=0,scale=0.9]{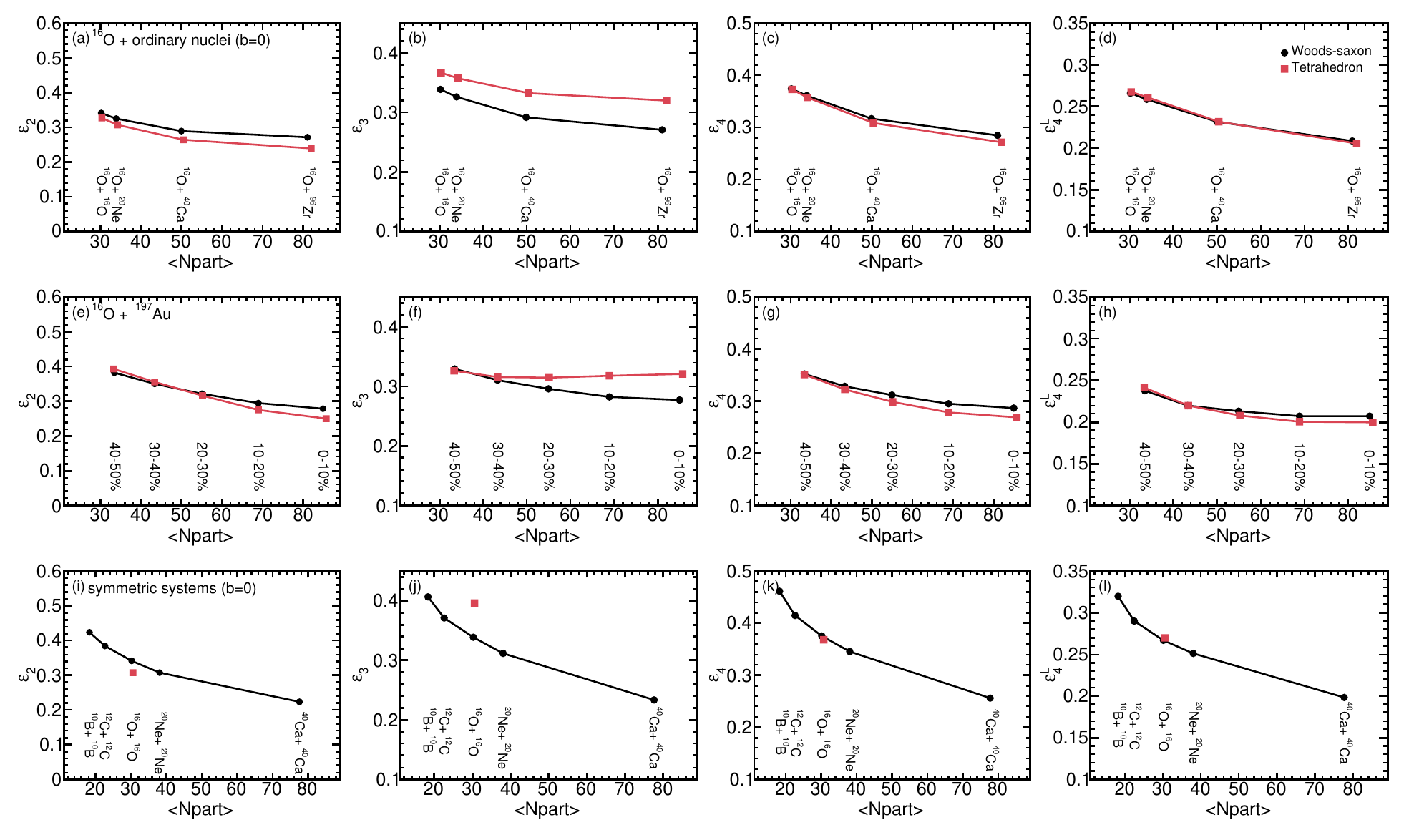}
	\caption{Eccentricity coefficients, namely $\varepsilon_2$, $\varepsilon_3$, $\varepsilon_4$ and $\varepsilon_4^L$ (from left column to right column) as a function of number of participants $\left<N_{part}\right>$. 
	Upper panels (Case I: (a), (b), (c), and (d)) are the results from  the most central collisions of the $\mathrm{^{16}O}$ nucleus (with or without $\alpha$-cluster) + ordinary nuclei ($^{16}$O, $^{20}$Ne, $^{40}$Ca and $^{96}$Zr), middle panels (Case II: (e), (f), (g), and (h)) are the results for the centrality dependence of the $\mathrm{^{16}O}$ (with or without $\alpha$-cluster) + $^{197}$Au collisions, and lower panels (Case III: (i), (j), (k), and (l)) represent the symmetric collision systems from small  to large ones in the most central collisions, i.e.  $^{10}$B + $^{10}$B, $^{12}$C + $^{12}$C, $^{16}$O +  $^{16}$O (with or without $\alpha$-cluster),  $^{20}$Ne +  $^{20}$Ne,  and  $^{40}$Ca + $^{40}$Ca.}
	\label{Fig1}
\end{figure*}

Eccentricities calculated by  Eq.~(\ref{eq:1})  is labeled as $\varepsilon _{2}$, $\varepsilon _{3}$ and $\varepsilon _{4}$, and linear part calculated by Eq.~(\ref{eq:4})  is labeled as $\varepsilon _{4}^{L}$ in the section of results and discussion. The eccentricity calculation results of different $\mathrm{\leftidx{^{16}}O}$ structures at $\sqrt{s_{NN}}$ = 6.37 TeV are plotted in Fig.~\ref{Fig1} and will be discussed later.

Usually  anisotropic flows can be characterized by the Fourier decomposition of the particle azimuthal distribution in the transverse plane through the equation
\begin{equation}
	\begin{aligned}
		\frac{dN}{d\varphi}\propto 1+2\sum_{n=1 }^{\infty }v_{n}cos[n(\varphi-\Psi _{n})]
		\label{eq:7}
	\end{aligned},
\end{equation}
here $\varphi$ and $\Psi_n$ are the azimuthal angles of final particles in momentum space and the event plane angle, respectively, $v_{n}$ is the $n$-th order flow coefficient. In the LHC experiments, there exists the phenomenon that the lower order anisotropic flow $v_{n}$ (n = 2, 3) is largely determined by a linear response to the corresponding $\varepsilon _{n}$ while higher order anisotropic flow have the contributions proportional to the product of $\varepsilon _{2}$ and/or $\varepsilon _{3}$. This can be easily understood because  each order anisotropic flow in the above equation has its own $n$-th order flow symmetric plane angle, the corresponding $cosin$ terms are not orthogonal with each other. That's to say, in principle, even for the lowest order anisotropic flows, they have the contributions from high order ones. Fortunately, according to the experiment findings, we only need to do the simplest decomposition as follows~\cite{YAN201582,PhysRevC92034903,201768},
\begin{equation}
	\begin{aligned}
		V_{4} = V_{4}^{NL}+V_{4}^{L} = \chi _{4,22}(V_{2})^{2}+V_{4}^{L}.
		\label{eq:8}
	\end{aligned}
\end{equation}

\begin{figure*}[htb]
	\includegraphics[angle=0,scale=0.9]{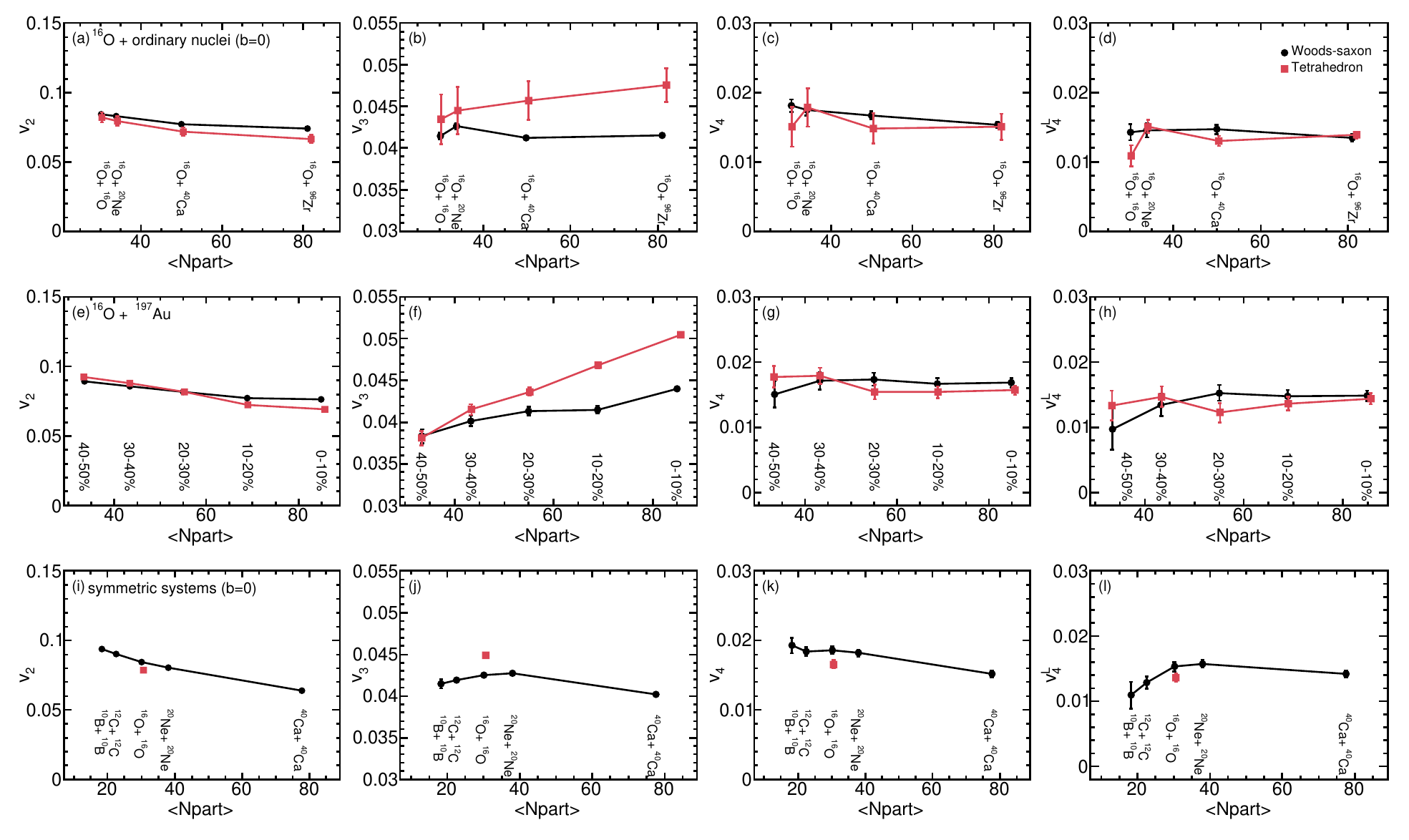}
	\caption{Same as Fig.~\ref{Fig1} but for anisotropic flow coefficients, namely $v_2$, $v_3$, $v_4$ and its linear mode  $v_4^L$.}
	\label{Fig2}
\end{figure*}

The complex flow vector here is defined as: $V_{n}\equiv v_{n}e^{in\Psi _{n}}$, where $v_{n}=\left | V_{n} \right |$ is the flow coefficient, and $\Psi _{n}$ represents the azimuth of $V_{n}$ in momentum space. The simplest approach to obtain $v_{n}$ is using the 2-particle correlations
\begin{equation}
	\begin{aligned}
		v_{n}\left \{ 2 \right \} = \left \langle \left \langle cosn(\varphi _{1}-\varphi _{2}) \right \rangle \right \rangle^{1/2}=\left \langle v_{n}^{2} \right \rangle^{1/2}
		\label{eq:10}
	\end{aligned}.
\end{equation}

To suppress non-flow effects, events are divided into two sub-events $A$ and $B$, separated by a pseudorapidity gap, and the equation is modified as
\begin{equation}
	\begin{aligned}
		v_{n}\left \{ 2 \right \} = \left \langle \left \langle cosn(\varphi _{1}^{A}-\varphi_{2}^{B}) \right \rangle \right \rangle^{1/2} = \left \langle v_{n}^{2} \right \rangle^{1/2}
		\label{eq:11}
	\end{aligned},
\end{equation}
\begin{equation}
	\begin{aligned}
		v_{4,22}^{A} = \frac{\left \langle \left \langle cos(4\varphi _{1}^{A}-2\varphi _{2}^{B}-2\varphi _{3}^{B}) \right \rangle \right \rangle}{\sqrt{\left \langle \left \langle cos(2\varphi _{1}^{A}+2\varphi _{2}^{A}-2\varphi _{3}^{B}-2\varphi _{4}^{B}) \right \rangle \right \rangle}}
		\label{eq:12}
	\end{aligned}.
\end{equation}

We can take the following approximation as the correlation between lower and higher flow coefficients is weak
\begin{equation}
\begin{aligned}
v_{4,22} = \frac{\left \langle v_{4}v_{2}^{2}cos(4\Psi _{4}-4\Psi _{2}) \right \rangle}{\sqrt{\left \langle v_{2}^{4} \right \rangle}}\approx \left \langle v_{4}cos(4\Psi _{4}-4\Psi _{2}) \right \rangle
\label{eq:13}
\end{aligned},
\end{equation}
and the magnitudes of the linear mode in higher order anisotropic flows is calculated as
\begin{equation}
\begin{aligned}
v_{4}^{L} = \sqrt{v_{4}^{2}\left \{ 2 \right \} - v_{4,22}^{2}}
\label{eq:14}
\end{aligned}.
\end{equation}

This acoustic scaling of linear and mode-coupled anisotropic flow calculations are put in Fig.~\ref{Fig2} in the Section of Results and discussion. To make sure that the visible qualitative results are not affected by the calculation method, we use another flow analysis with cumulants mentioned in Ref.~\cite{PhysRevC64054901} and find the qualitative conclusion is consistent.

\section{Results and discussion}
\label{sec:results}

To investigate exotic structure of $\alpha$-clustered $^{16}$O,  three sets of collision systems at center of mass energy $\sqrt{s_{NN}}$ = 6.37 TeV are considered in this work, namely case I:  the $\mathrm{^{16}O}$ nucleus (with or without $\alpha$-cluster) + ordinary target nuclei inside which nucleons are always in the Woods-Saxon distribution in most central collisions (b=0),  and these ordinary target nuclei include $^{16}$O, $^{20}$Ne, $^{40}$Ca and $^{96}$Zr; case II: the centrality dependence of $\mathrm{^{16}O}$ (with or without $\alpha$-cluster) + $^{197}$Au collisions, and case III: the symmetric collision systems from small systems to large ones, namely $^{10}$B + $^{10}$B, $^{12}$C + $^{12}$C, $^{16}$O +  $^{16}$O (with or without $\alpha$-cluster),  $^{20}$Ne +  $^{20}$Ne,  and  $^{40}$Ca + $^{40}$Ca,  in the most central collisions, in which nuclei except $^{16}$O are calculated with the ordinary Woods-Saxon nuclear structure.  
In this work, we define $\left<N_{part}\right>$ as the average number of participate nucleons in collisions, and for the case II we first determine the centralities by $N_{track}$ (number of final state charged particles)  and take the last five centralities (namely 40-50$\%$, 30-40$\%$, 20-30$\%$,10-20$\%$, 0-10$\%$) to calculate the $\left<N_{part}\right>$ for each centrality. The eccentricity coefficients were calculated through initial partons in the AMPT model and the anisotropic flow coefficients were analyzed including charged hadrons ($\pi ^{\pm }$, $K^{\pm }$, p and $\bar{p}$) with kinetic windows for the rapidity cut ($-0.5<y<0.5$) and transverse momentum cut ($0.2<p_{T}<3$) GeV/$c$.

 \begin{figure*}[htb]
	\includegraphics[angle=0,scale=0.9]{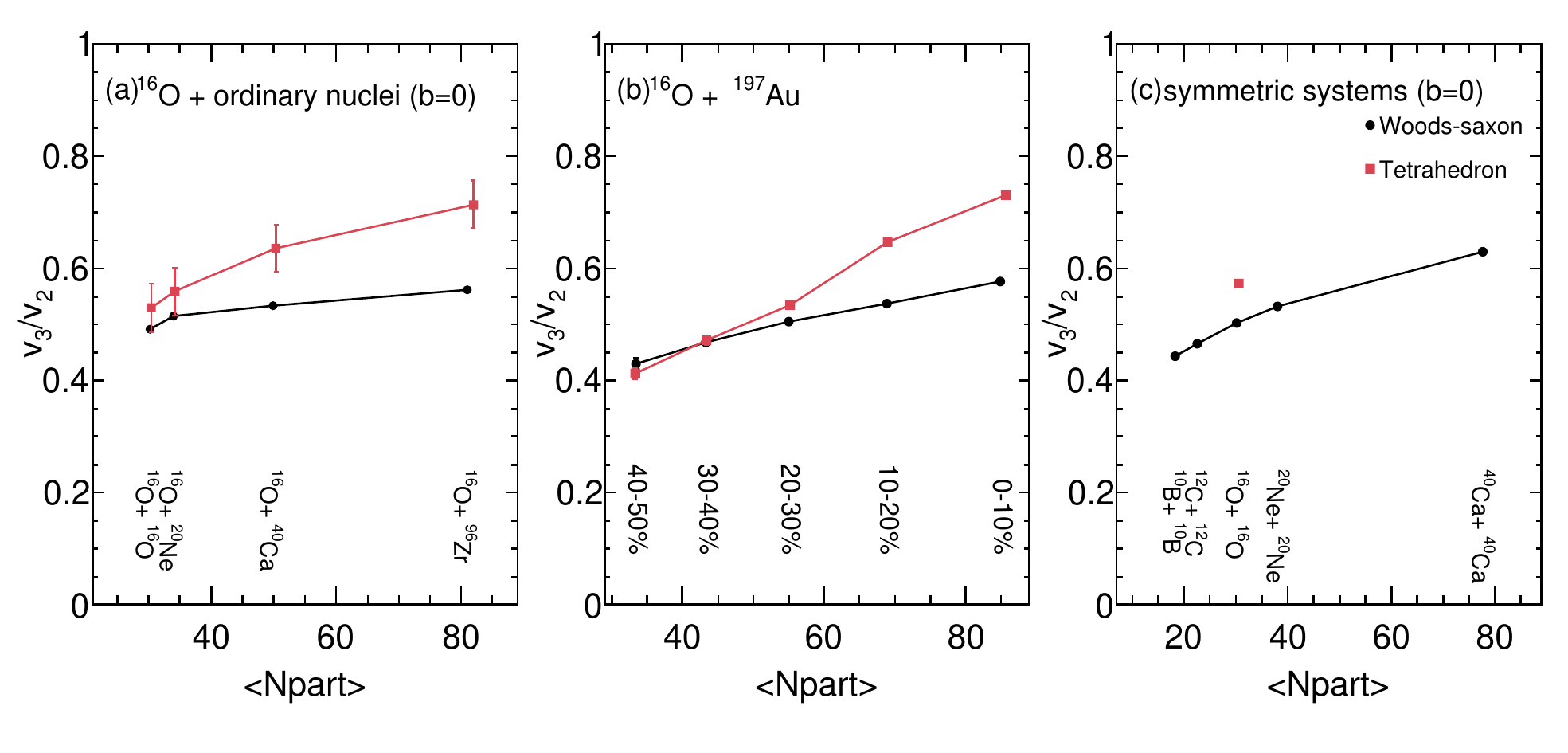}
	\caption{Ratio of $v_3/v_2$ as a function of $\left<N_{part}\right>$:  the case I, i.e.  $\mathrm{^{16}O}$  + ordinary nuclei at b = 0 fm (a), the case II,  i.e. $\mathrm{^{16}O}$ + Au at different centralities (b), the case III, i.e. symmetric collisions at b = 0 fm (c). The red or black lines (symbols) represent $^{16}$O w/ or w/o $\alpha$-cluster structure.}
	\label{Fig3}
\end{figure*}

 \begin{figure*}[htb]
	\includegraphics[angle=0,scale=0.9]{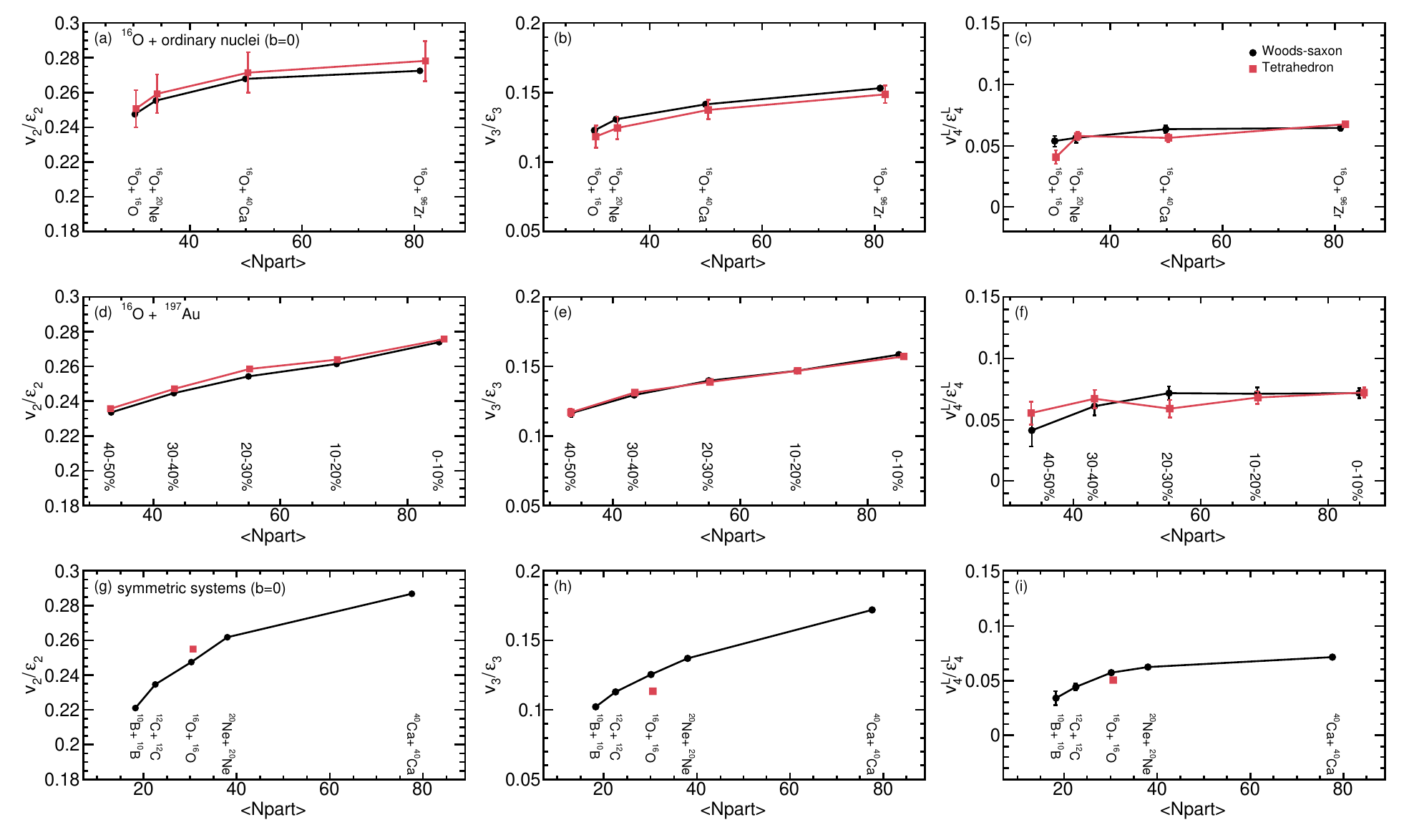}
	\caption{Ratios of $v_2/\varepsilon_2$ (left column) ,  $v_3/\varepsilon_3$ (middle column) and $v_4^L$/$\varepsilon_4^L$ (right column)  as a function of number of participant $\left<N_{part}\right>$: The case I (upper row),  i.e.  $\mathrm{^{16}O}$  + ordinary nuclei at b = 0 fm (a, b,  c);  the case II (middle row), i.e. $\mathrm{^{16}O}$ + Au at different centralities (d, e, f),  and the case III (bottom row), i.e. symmetric collisions at b = 0 fm  (g, h, i).  The red or black lines (symbols) represent $^{16}$O w/ or w/o $\alpha$-cluster structure.}
	\label{Fig4}
\end{figure*}

Figure ~\ref{Fig1} shows  the eccentricity coefficients $\varepsilon _{n}$ ($n$ = 2, 3, 4) and $\varepsilon _{4}^{L}$ of different collision systems (labeled with $\left<N_{part}\right>$) at center of mass energy $\sqrt{s_{NN}}$ = 6.37 TeV  from left to right columns.
Panel (a), (b), (c) and (d) present $\varepsilon _{n}$ ($n$ = 2, 3, 4) and $\varepsilon _{4}^{L}$ for the case mentioned above, namely different structured $\mathrm{^{16}O}$ nucleus + ordinary nuclei ($\mathrm{\leftidx{^{16}}O}$, $\mathrm{\leftidx{^{20}}Ne}$, $\mathrm{\leftidx{^{40}}Ca}$, $\mathrm{\leftidx{^{96}}Zr}$ always have the Woods-Saxon nucleon distribution) in the most central collisions. In this case, $\varepsilon _{2}$ decreases with the increasing of the collisions system size and tetrahedron structure configuration presents a little higher value of $\varepsilon _{2}$.
$\varepsilon _{4}$ gives the similar system size dependence of $\varepsilon _{2}$ and $\varepsilon _{4}^L$ take approximately the same value between the patterns with different configurations of $\mathrm{^{16}O}$. $\varepsilon _{3}$ also shows the decreasing trend with system size, and in each given collision system $\mathrm{^{16}O}$ in tetrahedron structure gives larger $\varepsilon _{3}$ than that of the Woods-Saxon structure.
Panel (e), (f), (g), and  (h) in Fig.~\ref{Fig1} displays centrality ($\left<N_{part}\right>$) dependence of $\varepsilon _{n}$ ($n$ = 2,3,4) and $\varepsilon _{4}^{L}$ in $\mathrm{^{16}O}$ + $\mathrm{^{197}Au}$ collisions, i.e. the case II mentioned above. It is obviously that all $\varepsilon _{2}$, 
$\varepsilon _{3}$ and $\varepsilon _{4}$ decrease with the increasing of $\left<N_{part}\right>$ except for the $\varepsilon _{3}$ with configuration of $\mathrm{^{16}O}$ in tetrahedron structure. Tetrahedron structure configuration of $\mathrm{^{16}O}$ also presented smaller $\varepsilon _{2}$ ($\varepsilon _{4}$) and larger $\varepsilon _{3}$ than the Woods-Saxon configuration did in this case.
For case III, $\varepsilon _{n}$ (n=2,3,4) and $\varepsilon _{4}^{L}$ as a function of $\left<N_{part}\right>$ present a downward trend for the configuration of initial nucleon in the Woods-Saxon distribution, as shown in panel (i), (j), (k) and (l), respectively. The assumed $\alpha$-clustered $\mathrm{^{16}O}$ in tetrahedron gives an obvious deviation for the eccentricity coefficients $\varepsilon _{3}$ than that of the Woods-Saxon configuration.
We can see that $\varepsilon _{4}$ takes the order of $\varepsilon _{2}$ and if we subtract the contributions of $\varepsilon _{2}$, namely $\varepsilon _{4}^{L}$, the results are almost the same for different $^{16}$O structures. Here we have the first conclusion: In the above collision systems, quite large part of $\varepsilon _{4}$ is originated from the $\varepsilon _{2}$ especially for small systems, which is important for the understanding of the flow calculations shown below. 
We end here and do not show higher-order calculation results because the figure shows the higher-order calculation is somewhat unnecessary  in small systems due to the fluctuations.

Eccentricity coefficients reflect both initial geometry distribution and initial fluctuations. In this context, the system scan project could be a potential way to distinguish the initial intrinsic geometry distribution, depending on how sensitive the final observables in momentum space, such as anisotropic flows, to the initial geometry distribution.

Figure~\ref{Fig2} shows the anisotropic flow coefficients $v_n$ ($n$ = 2, 3, 4) and the fourth order linear mode  $v_4^L$ for the case I, i.e. $\mathrm{^{16}O}$  + ordinary nuclei at b = 0 fm, which are presented in panels (a), (b), (c), and (d); the case II, i.e. $\mathrm{^{16}O}$ + Au at different centralities,  which are presented in panels in (e), (f), (g), and (h); and the case III, i.e. various symmetric collisions at b = 0 fm,  which are presented in (i), (j), (k), and (l). We find that anisotropic flow coefficients present the similar $\left<N_{part}\right>$ dependence as the eccentricity coefficients. Since $v_4$ shows similar $\left<N_{part}\right>$ dependence with $v_2$, we can see this in panels (c), (g), (k),  and (a), (e), (i), $v_2$ and $v_3$ will be discussed for distinguishing the initial geometry distribution. 
$v_n$ ($n$ = 2, 3, 4)  from the configuration of $\mathrm{\leftidx{^{16}}O}$ in tetrahedron 4-$\alpha$ structures are obviously deviated from the Woods-Saxon configuration as eccentricity discussed above in the case III. From the case III, the Woods-Saxon configuration presents a smooth $\left<N_{part}\right>$ dependence of anisotropic flow (or eccentricity), and at the point of O + O collisions, there occurs a peak or dip if $\mathrm{\leftidx{^{16}}O}$ is in tetrahedron $\alpha$-clustered structure. 
From the case I and II, the $\left<N_{part}\right>$ dependence trend of $v_3$ can also distinguish the tetrahedron $\alpha$-clustered $\mathrm{\leftidx{^{16}}O}$ from the Woods-Saxon distribution. Note that the collective flow coefficients in $^{16}$O+$^{16}$O collisions were consistent with the theoretical works~\cite{PhysRevC99044904} by using the AMPT model with nuclei of the Woods-Saxon configuration.

To further compare three cases, the ratio of anisotropic flow coefficients $v_3/v_2$ was presented in Fig.~\ref{Fig3}. For the case I, $v_3/v_2$ approximately keeps flat as a function of $\left<N_{part}\right>$ for configurations of $\mathrm{\leftidx{^{16}}O}$ in the Woods-Saxon distribution, however, it increases with $\left<N_{part}\right>$ for the $\alpha$-clustered tetrahedron $\mathrm{\leftidx{^{16}}O}$ structure. For the case II, the ratio of $v_3/v_2$ presents an upward trend with $\left<N_{part}\right>$ in the Woods-Saxon distribution as well as the tetrahedron configuration of $\mathrm{\leftidx{^{16}}O}$. From the case I and II, the collision system as a ``magnifier" can enlarge the ratio of $v_3/v_2$ with the increasing of the system size. 
For the case III, $v_3/v_2$ displays a $\left<N_{part}\right>$ dependence  for the Woods-Saxon distribution, and the tetrahedron configuration of $\mathrm{\leftidx{^{16}}O}$ results in an enhanced point  beyond the Woods-Saxon baseline. The source of initial geometry for the case II, namely centrality dependence in O + Au collisions, contains more complex components, such as nuclear intrinsic geometry, initial fluctuation and geometry of overlap region between target and projectile nuclei. Cases I and III are all chosen in the most central collisions, which can also be achievable in experiment, to avoid the geometry distribution from overlap region as much as possible. And from the above results and discussion, it is concluded that $v_3/v_2$ can be taken as a probe to identify $\alpha$-clustered structure of $\mathrm{\leftidx{^{16}}O}$, and the cases I and III are proposed as a potential scenario  of system scan experiment project at RHIC or LHC.

 From the propositions of hydrodynamics~\cite{SongNST,HydroKN-1,HydroKN-2,PhysRevC86044908,PLB2015_nonlinear2,vnen_beta_Ntrack201808}, the relationship between initial geometry and final anisotropic flow can be described by $v_n \propto \varepsilon_n$ for lower orders $n$ = 2, 3 and $v_n^L$ $\propto$ $\varepsilon_n^L$ for higher orders $n>3$. These relations provide efficiency information of the transformation from initial geometry properties to final momentum space in heavy-ion collisions. 
 Figure~\ref{Fig4} shows the ratio of $v_n/\varepsilon_n$ ($n$ = 2,3) as well as  $v_4^L/\varepsilon_4^L$ for the case I: panels (a), (b), and (c); the case II: panels (d), (e), and (f); and the case III: panels (g), (h), and (i). All ratios increase with $\left<N_{part}\right>$ and show no significant difference between two configurations of $\mathrm{\leftidx{^{16}}O}$. This implies the transformation efficiency is quite similar for these collision systems, and both ratios of $v_n/\varepsilon_n$ ($n$ = 2, 3) and $v_4^L/\varepsilon_4^L$ seem to only depend on the system size, such as $\left<N_{part}\right>$ at a given center of mass energy.  It is noted that the values of  $v_n/\varepsilon_n$ ($n$ = 2, 3) and $v_4^L/\varepsilon_4^L$ are related to the ratio of shear viscosity ($\eta$) over entropy density ($s$) of hot-dense matter as pointed in some previous studies ~\cite{vnen_beta_Ntrack201808,ZHANG2020135366}, 
the fact that  the insensitivity to geometrical  configuration of $v_2/\varepsilon_2$, $v_3/\varepsilon_3$ and $v_4^L/\varepsilon_4^L$ in the present calculation in turn provides us a possibility to extract $\eta/s$ if the suitable viscous hydrodynamics model is used, which also indicates that $\eta/s$ might be insensitive to  the initial geometric structure. 
 On the other hand, due to the cancel-out effect of $\eta/s$,   the $v_3/v_2$  shall be a good probe to identify the geometric structure regardless of the $\eta/s$.   

Finally, to give an illustrative interpretation for $v_3$'s sensitivity to the geometric structure, we consider to project the  $\mathrm{^{16}O}$ nucleus into the transverse plane after a 3D rotation, there will be a larger probability to see some projected images due to  its highly structural symmetry. Take the tetrahedron structure for an example, if we draw the projection of participated nucleons in the initial collision or the density distribution of partons in the HIJING procedure, it is more likely to observe triangular images than that of the Woods-Saxon structure, which results in large triangularity flow $v_{3}$. Besides that, fluctuation in small system also plays an important role in the final state. As increasing of the system size, fluctuation becomes weaker and intrinsic geometry will contribute more to the eccentricity coefficients and then the final collective flow. Therefore the collision system dependence of final observables which are sensitive to initial geometry properties, can indicate the intrinsic geometry distribution and then can be taken as a probe to distinguish the $\alpha$-clustering nuclear structure.

\section{summary}
In summary, the present study shows the AMPT calculations of anisotropic flows in relativistic heavy ion collisions including $\mathrm{^{16}O}$  
which is assumed having exotic tetrahedral  structure with 4 $\alpha$-clusters.  Three different sets of system scan at center of mass energy $\sqrt{s_{NN}}$ = 6.37 TeV were considered.  The case I is the ordinary structured target size scan by the different configured $\mathrm{^{16}O}$ projectile in  the most central collisions, the case II presents the centrality scan for  $\mathrm{^{16}O}$ + $\mathrm{^{197}Au}$ collisions, and the case III describes  the symmetric collision system scan  (namely B + B, C + C, O + O, Ne + Ne, and Ca + Ca) in the most central collisions. 
From the systematic calculation results of the above three cases, it demonstrates that $v_{3}$, $v_{2}$ and the ratio of $v_{3}/v_{2}$ can be taken as a powerful probe to distinguish the tetrahedral configuration of $\mathrm{\leftidx{^{16}}O}$ from the Woods-Saxon  configuration in the ground state. And the collision systems for the case I and the case III are proposed to be the system scan experiment projects at RHIC or LHC.

\begin{acknowledgements}

This work was supported in part by the National Natural Science Foundation of China under Contracts No. 11890714, 11875066, 11421505 and 11961141003, 11935001 and the National Key R\&D Program of China under Grants No. 2016YFE0100900 and 2018YFE0104600, the Strategic Priority Research Program of the CAS under
Grants No. XDB34030200  and XDB16, and the Key Research Program of Frontier Sciences of the CAS under Grant No. QYZDJ-SSW-SLH002.
\end{acknowledgements}

\end{CJK*}	
		\bibliography{no}

\end{document}